# Deep-subwavelength features of photonic skyrmions in a confined electromagnetic field with orbital angular momentum

Luping Du[1,*], Aiping Yang[1], Anatoly V. Zayats[2,*] and Xiaocong Yuan,[1,*]

[1]Nanophotonics Research Centre, Shenzhen Key Laboratory of Micro-Scale Optical Information Technology, Shenzhen University, 518060, China

[2]Department of Physics, King's College London, Strand, London, WC2R 2LS, United Kingdom

[*]Correspondence: lpdu@szu.edu.cn, xcyuan@szu.edu.cn, and a.zayats@kcl.ac.uk

In magnetic materials, skyrmions are nanoscale regions where the orientation of electron spin changes in a vortex-type manner. Here we show that spin-orbit coupling in a focused vector beam results in a skyrmion-like photonic spin distribution of the excited waveguided fields. While diffraction limits the spatial size of intensity distributions, the direction of the field, defining photonic spin, is not subject to this limitation. We demonstrate that the skyrmion spin structure varies on the deep-subwavelength scales down to 1/60 of light wavelength, which corresponds to about 10 nanometre lengthscale. The application of photonic skyrmions may range from high-resolution imaging and precision metrology to quantum technologies and data storage where the spin structure of the field, not its intensity, can be applied to achieve deep-subwavelength optical patterns.

Electromagnetic waves can carry both spin (determined by their polarization) and orbital angular momenta [1,2]. These are considered as two separate degrees of freedom of a paraxial beam [3]. However, spin-orbit coupling takes place in focused beams or evanescent waves [4-6], laying the foundations for a variety of applications in physics, optics, and metrology [7,8]. Many analogies have recently been developed between electron and optical spin effects, resulting in innovative applications in metrology, sensing, optical information and quantum technologies [3].

In magnetic materials, one of the prominent manifestations of spin-orbit coupling is the appearance of skyrmions—nanoscale vortexes of electron spins [9-11]. Here we show that spin-orbit coupling in a focused vector beam results in a skyrmion-like photonic spin distribution with the photon spin vector either along or opposite to orbital angular momentum, depending on the spatial location in a beam cross-section. For waveguided modes with evanescent fields, this gives rise to a Neel-type photonic skyrmions; while a Bloch-type skyrmion-like photonic spin structure is formed within the central area of focused free-space propagating beams. While diffraction phenomenon limits the size of spatial intensity distributions determined by the electromagnetic field amplitude, the direction of the field, defining its polarization, is not subject to this limitation and can be designed at much finer, deep-subwavelength scales. We demonstrate that the spin structure of a focused vortex beam varies on the deep-subwavelength scales down to 1/60 of light wavelength, which corresponds to about 10 nanometre lengthscale. We show that such subwavelength spin variations can be readily designed since the presence of a spiral phase in the axial field of the beam carrying orbital angular momentum intrinsically accompanies the spin associated with the transverse field of the beam. The application of photonic skyrmions may range from high-resolution imaging and precision metrology to quantum technologies and data storage where the analysis of the spin structure of the beam, not its intensity, can be applied to achieve deep-subwavelength optical field patterns.

All waveguided and surface electromagnetic waves have an evanescent field component and carry spin angular momentum (SAM) which is oriented perpendicular to the direction of their propagation, termed a transverse spin [12]. Optical vortex beams with helical wavefronts also carry intrinsic orbital angular momentum (OAM) described by the vortex topological charge ($L$) which determines the phase increment around the vortex core [13]. The intensity and spin structure of the evanescent vortex beam with a topological charge $L = +1$ is shown in Fig. 1 (see Supplementary Information for the details of the simulations and Fig. S1). The spin-orbit coupling in such a beam results in a fine structure of the spin state and leads to appearance of a longitudinal SAM ($S_z$), not present in the evanescent waves without vortices: the oscillations of the electric field vector in a plane perpendicular to the propagation direction (transverse electromagnetic field) can be described by a spin vector parallel to the propagation direction. These longitudinal ($S_z$) and transverse ($S_r$) SAMs of the evanescent vortex beam vary across the beam crossection and form an optical spin texture (Fig. 1a,b). A progressive change of the spin vector is seen from the "up" state in the centre to the opposite, "down" state at the position $r=r_0$ along the radial direction. This photonic spin structure of the evanescent vortex beam can be proved to be a direct analogue with the structure of the magnetisation direction in the Neel-type skyrmions (skyrmion number n=1, see Supplementary Materials) in magnetic materials [11]. It is interesting to note that the free-space propagating nondiffracting vortex vector beams carry the longitudinal and azimuthal SAM components and do not exhibit skyrmion-like spin structure; at the same time, free-space focused propagating vortex beams have a spin structure which can be approximated by a Bloch-type skyrmion, in contrast to Neel-type skyrmions in the evanescent vortex beams (see Supplementary Information).

We experimentally demonstrated the spins structure of the evanescent vortex beams on the example of plasmonic vortices [14,15]. We developed a unique scanning near-field optical microscopy set-up which works in a dark-field-like configuration in order to extract a weak Rayleigh scattering signal from the near-field probe (a polystyrene nanoparticle in this work) from the strong background illumination, enabling super-resolved characterisation of the evanescent field with high signal-to-noise ratio (See SI for the details of the experimental set up). The experimental maps of the spin structure of the plasmonic vortex beams with angular momenta $L$ = +1, 0 and -1 are presented in Fig. 2. The intensity distributions of RCP and LCP light generated by scattering of the SPP vortex beam is in line with the theoretical predictions (Fig. 1 a,c and Fig. S5). Both the experiment and simulations show that across the beam cross-section, the local spin of the in-plane electric field varies dramatically with the distance to the beam axis. The obtained spin structures for the plasmonic vortices with $L$ = ±1 reveal the multiple reversals of the spin state across the vortex, with the spin state in the centre of the beam determined by the net angular momentum of the excitation beam. For a radially polarized incident beam [16] ($L$ = 0), the spin structure is not observed, as expected because of the absence of OAM associated with the excited surface plasmon polaritons (SPPs). For the beams with $L$ = ±1, the spin reversal caused by OAM-SAM coupling is clearly observed with the spin achieving opposite values to the spin of the incident beam, showing efficient OAM in SAM conversion.

In order to characterize the longitudinal spin, we define a spin-related parameter $\gamma_s=(I_{RCP}-I_{LCP})/(I_{RCP}+I_{LCP})\propto S_z/I$, where $I_{RCP}$ and $I_{LCP}$ are the intensities of the right- and left-handed circularly polarized components of the transverse field, respectively, and $I= I_{RCP}+I_{LCP}$. Thus, $\gamma_s$ = +1(-1) represents a pure right (left) -handed circular polarization, $\gamma_s$=0 represents a linear polarization, and fractional values represent the elliptical polarizations.

In the evanescent vortex beam, the first spin reversal is followed by multiple flips of increasing frequency with the increasing distance (Figs. 1, 3 and Fig. S5). The distance required to complete a spin reversal from $\gamma_s$ = +1 to $\gamma_s$ = -1 state is plotted in Fig. 1d. Except for the first reversal close to the beam axis for which the distance is on the order of the diffraction limit, the other spin reversals takes place at the deep-subwavelength scales. For example, the second spin reversal requires only approximately $\lambda/17$ distance. While the intensity variations within a light beam are subject to diffraction limit (~ $\lambda/2$), the spin structure which is determined by the direction and not the amplitude of the electric field is not governed by diffraction and can achieve deep-subwavelength resolution. This provides a new concept to modulate optical fields on the deep-subwavelength scales using focused vortex beams which can be used for applications in precision metrology, ultra-sensitive displacement sensor, super-resolution imaging, chiral molecule detection, to name but a few. In the above theoretical analysis, the only assumption is a spiral phase associated with the axial field, so the considerations are applicable to any type of vortex beams with this property, being free-space, evanescent or waveguided modes.

The high-resolution cross-sectional profile of the spin state reversal is shown in Fig. 3a for the $L = +1$ beam. It reveals the changes of the spin state from positive to negative on subwavelength scales. Away from the beam centre, the second spin reversal (Fig. 3c, light green area in Fig. 3a) takes place over the distance of 53 nm (37.5 nm from the simulations, Fig. 3d). A more interesting for practical applications is the full-width at half-maximum (FWHM) of the sharp dip between the peaks, which is approximately 15 nm ($\lambda/45$) measured in the experiment (approximately 10 nm or $\lambda/63$ from simulations). The deviation of the experimental measurements from the theoretical prediction is probably caused by the size of the probe nanoparticle which limits the spatial resolution. These observed sharp features of optical fields, experimentally demonstrated here on the example of SPPs but characteristic for all kinds of waveguided modes and focused vortex beams, can be engineered using spin-orbit coupling in focused vortex beams and serve as a probe for deep-subwavelength imaging, high precision metrology and chirality mapping.

In summary, an intrinsic connection and transformations between the two forms of angular momenta associated with a confined electromagnetic field allows engineering of local spin in the focused beam irrespectively of the spin of the incident light. The observed spin variations resemble skyrmion structure associated with magnetization orientation variations in magnetic materials and can be considered as a photonic analogue of magnetic skyrmions, bringing about exciting applications in metrology, sensing, data storage and quantum technologies.

**REFERENCES.**

**Acknowledgments.** L. Du thanks S. Peng for his assistance with the theoretical analysis. This work was supported, in part, by the National Natural Science Foundation of China grants 61622504, 61427819, 61490712 and 11504244, National Key Basic Research Program of China (973) grant 2015CB352004, the leading talents of Guangdong province program grant 00201505, the Natural Science Foundation of Guangdong Province grant 2016A030312010, the Science and Technology Innovation Commission of Shenzhen grants KQTD2015071016560101, KQTD2017033011044403, ZDSYS201703031605029, EPSRC (UK) and ERC. A.Z. acknowledges support from the Royal Society and the Wolfson Foundation.

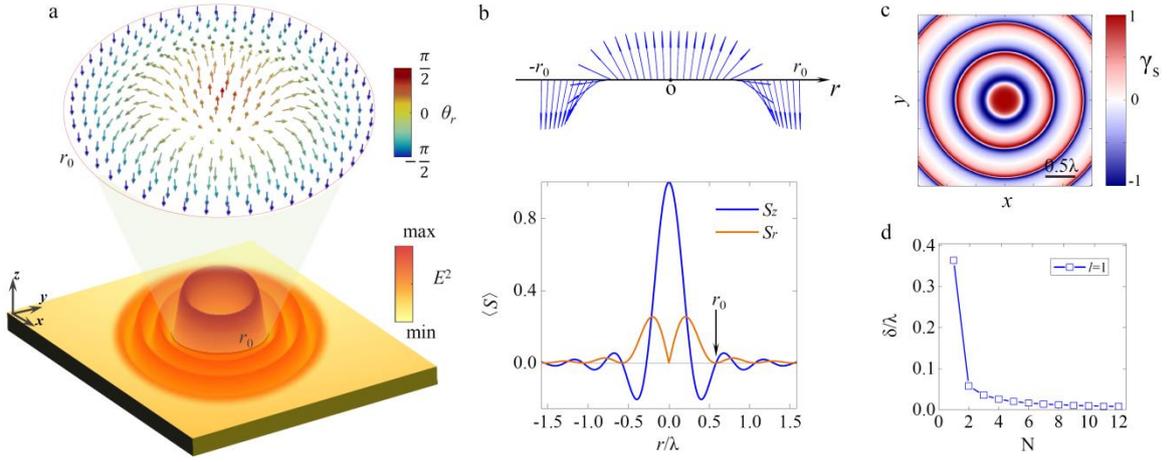

Figure 1. Spin structure of the evanescent field vortex forming the Neel-type photonic skyrmion. (a) Intensity distribution of the evanescent field vortex with a topological charge $L=1$ (bottom) and the distribution of the photonic spin orientation in the centre of the vortex (top). The arrows indicate the direction of the unit spin vector determined by $\vartheta_r=\arctan(S_z/S_r)$. (b) The radial variations of the transverse ($<S_r>$) and longitudinal ($<S_z>$) spin components of the beam (bottom) and the orientation of the spin vector in the centre of the beam (top). (c) The simulated spin distribution (in term of $\gamma_s=(I_{RCP}-I_{LCP})/(I_{RCP}+I_{LCP})$) in the evanescent field in (a). (d) The plot of spatial distance ($\delta$) that is required to finish a spin reversal from a positive state ($\gamma_s=+1$) to the negative ($\gamma_s=-1$) across the radial direction. The in-plane wavevector ($k_r$) of the evanescent vortex for the above calculation is set to $1.05k_0$, where $k_0=2\pi/\lambda$ represents the wavevector in the free space.

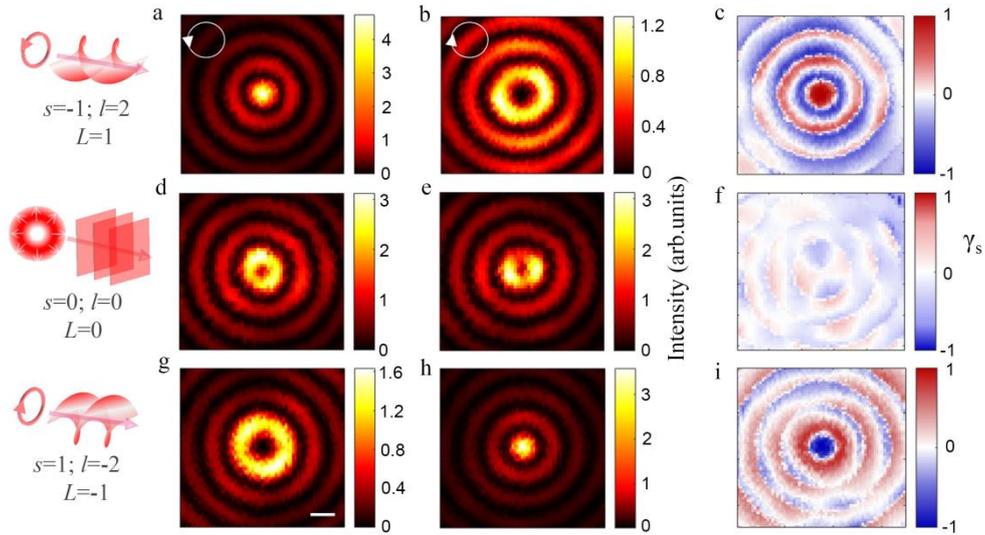

Figure 2. Spin-structure of a plasmonic vortex. (a-h) The intensity distributions of (a,d,g) RCP and (b,e,h) LCP components of light scattered by a nanoparticle scanning across a plasmonic vortex, and (c,f,i) the corresponding spin structure of the plasmonic vortex calculated as $\gamma_s=(I_{RCP}-I_{LCP})/(I_{RCP}+I_{LCP})$. Spin structure of the plasmonic vortices with topological charge of $L=+1$ (a,b,c), $L=0$ (d,e,f) and $L=-1$ (g,h,i) were mapped. The residual spin structure for $L=0$ in (f) arises mainly from the effect of a linear polarizer, which results in a slightly elliptical mapped intensity profile which rotates with the polarizer rotation. The SPP were excited on a surface of a 50 nm thick Ag film by light with a wavelength of $\lambda = 632.8$ nm. A polystyrene nanoparticle of a diameter of 320 nm was used as a scatterer. Scanning step size is 30 nm. All images are of the same size; the scale bar in (g) is $\lambda/2$. The inserts on the left show the structure of the field of the optical beams with different spin ($s$) and orbital ($l$) angular momenta leading to the excitation of plasmonic vortices with $L=s+l$. Simulated maps for the same beam parameters are shown in Extended Data Fig. 5.

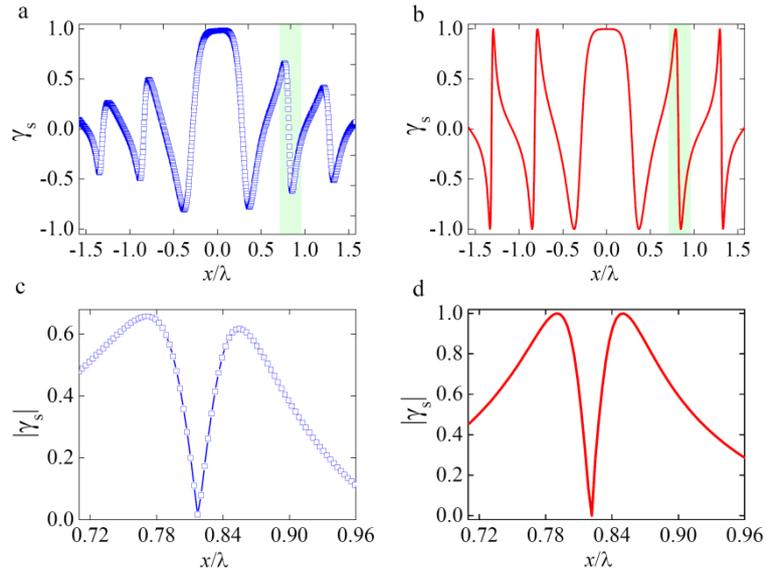

Figure 3. Characterization of the spin reversal in the photonic skyrmion. (a) Measured and (b) simulated traces of the spin state near the centre of a plasmonic vortex generated with *L*=+1 beam. (c,d) The detailed view of the spin structure of the areas indicated with green boxes in (a,c), where the spin reversal takes place on a deep-subwavelength scale. The absolute value of the spin is shown to facilitate the quantitative characterization of the feature size. In (a,c), a scanning step is 2 nm. All other experimental parameters are as in Fig. 2.

# Supplementary Information

# Deep-subwavelength features of photonic skyrmions in a confined electromagnetic field with orbital angular momentum

1. **Experimental setup for characterizing the spin structure in a surface plasmon vortex.**

The experimental setup for studies of the super-fine spin structure of a tightly-focused surface plasmon polariton (TM-type evanescent wave) on a metal-air interface is shown in Fig. S3. An incident laser beam (a wavelength of 632.8 nm) is tightly focused using an oil-immersion objective lens (Olympus, 100×, NA=1.49) onto the sample consisting of a 3-layer structure of a 50 nm thick silver film on a silica substrate with air as a superstrate. This provides an excitation of SPPs on an air-silver interface. A polystyrene (PS) nanoparticle of a diameter of 320 nm is immobilized onto the silver surface using 4-MBA molecular linkers in order to scatter the SPP in a far-field. The scattered light which contains the local spin information is collected by another objective lens (Olympus, 60×, NA=0.7) and sent through a beam splitter to 2 polarization analysing systems each composed of a combination of $\lambda/4$ waveplate and linear polarizer. One set is arranged at an angle of $45^o$ between the waveplate and the polarizer to detect only RCP light and the other set at an angle of $-45^o$ to detect the LCP light. After polarization selection, the intensity of light of each CP components is measured with photomultipliers (PMT). The sample is fixed onto a piezo-scanning stage (resolution of 1 nm) for mapping the spin structure by scanning the nanoparticle across the beam profile.

A typical back focal image of the reflected beam from the 3-layer structure is captured with CCD1 and is shown in Fig. S4 (a), where a dark ring indicates the excitation of the SPPs at the air-silver interface. In order to spatially separate a weak SPP scattering from a nanoparticle and a directly transmitted background light, a circular opaque mask is inserted into the beam path and an annular beam is formed to illuminate a sample via the focusing objective lens. The size of the mask is carefully designed in order not to affect the excitation of the SPPs while to block incident light at low incident angles (Fig. S4 (b)). The annular beam is over-focused slightly onto the sample such that the SPP scattering from a nanoparticle is collected and focused at a point different from the transmitted annular beam along the optical axis. The Rayleigh scattering image of a single PS nanosphere captured at CCD2 is shown in Fig. S4 (c), from which we can clearly observe the spatial separation of the Raleigh scattering signal from the particle and the transmitted light. As a result, by placing a fiber coupler (which acts as a pinhole) near the imaging plane of the collection objective lens, the scattered light can be separated from the background direct transmission. Fig. S4 (d) shows a dark field image of the PS nanoparticles immobilized on the silver film, demonstrating well separated particles allowing for single-nanoparticle studies.

## 2. Calculations of electromagnetic field in an optical system with axial symmetry.

In order to illustrate the spin-orbit properties in a confined electromagnetic field in an axisymmetric optical system, we consider a propagation of a beam in a source free, homogeneous and isotropic medium for which the Hertz vector potential ($\Pi$) can be introduced to reduce the Maxwell's equations to a single scalar differential equation [22]. Assuming exp($-i\omega t$) time dependence and propagating in $z$-direction (exp($ik_z z$)), the electromagnetic fields of a transverse magnetic (TM) mode in a cylindrical coordinate system ($r, \varphi, z$) can be described as

$$\boldsymbol{E} = [E_r, \ E_\varphi, \ E_z] = \left[ ik_z \frac{\partial}{\partial r}, \ i\frac{k_z}{r}\frac{\partial}{\partial \varphi}, \ k_r^2 \right] \Pi$$

$$\boldsymbol{H} = [H_r, \ H_\varphi, \ H_z] = \left[ -\frac{i\omega\varepsilon}{r}\frac{\partial}{\partial \varphi}, \ i\omega\varepsilon \frac{\partial}{\partial r}, \ 0 \right] \Pi \tag{1}$$

where $\varepsilon$ is the permittivity of the medium, $\omega$ is the angular frequency of the wave, $k_r$ and $k_z$ are the in-plane and axial wave-vector components, respectively, obeying the relation $k_r^2+k_z^2=k^2=\varepsilon_r k_0^2$, $\varepsilon_r=\varepsilon/\varepsilon_0$ is the relative permittivity of the medium, and $k_0$ and $k$ are the wavevectors in free space and in the medium, respectively. Similarly, transverse electric (TE mode) can be described as

$$\boldsymbol{H} = [H_r, \ H_\varphi, \ H_z] = \left[ ik_z \frac{\partial}{\partial r}, \ i\frac{k_z}{r}\frac{\partial}{\partial \varphi}, \ k_r^2 \right] \Pi$$

$$\boldsymbol{E} = [E_r, \ E_\varphi, \ E_z] = \left[ \frac{i\omega\mu}{r}\frac{\partial}{\partial \varphi}, \ -i\omega\mu \frac{\partial}{\partial r}, \ 0 \right] \Pi \tag{2}$$

where $\mu=1$ is the permeability of the medium. Since both TE and TM modes possess a similar spin property, we only consider TM-modes in the following calculations and analysis.

The Hertz vector potential satisfies the following Hertz wave equation in the cylindrical coordinate system:

$$\frac{1}{r}\frac{\partial}{\partial r}\left( r \frac{\partial \Pi}{\partial r} \right) + \frac{1}{r}\frac{\partial}{\partial \varphi}\left( \frac{1}{r}\frac{\partial \Pi}{\partial \varphi} \right) + k_r^2 \Pi = 0 \tag{3}$$

with

$$\Pi(r,\varphi,z) = AJ_l(k_r r)\exp(il\varphi)\exp(ik_z z) , \tag{4}$$

where *A* is a complex-valued constant and $J_l$ denotes the Bessel function of the first kind of the order *l*. This Herz vector potential describes the electromagnetic field with a spiral phase of a topological charge "*l*" (20). Subsequently, each components of electric and magnetic fields can be obtained using Eqs. (1,2).

## 3. Theoretical analysis of photonic spin structure in an evanescent optical vortex.

We first consider the situation for an evanescent optical vortex (e-OV). In this case, the in-plane wave-vector component ($k_r$) is large than the wave-vector (*k*) of beam, resulting in an imaginary axial wave-vector component (denoted as $ik_z$). By replacing the z-component wave-vector in Eq. (1) and Eq. (4) with the imaginary value: $ik_z$, we can obtain the electromagnetic fields for a e-OV of a TM-polarised field as

$$\boldsymbol{E} = \begin{pmatrix} E_r \\ E_\varphi \\ E_z \end{pmatrix} = \begin{pmatrix} -Ak_z k_r J_l'(k_r r) \\ -Ai \dfrac{lk_z}{r} J_l(k_r r) \\ Ak_r^2 J_l(k_r r) \end{pmatrix} e^{il\varphi} e^{-k_z z}$$

$$\boldsymbol{H} = \begin{pmatrix} H_r \\ H_\varphi \\ H_z \end{pmatrix} = \begin{pmatrix} A\dfrac{\omega\varepsilon l}{r} J_l(k_r r) \\ Ai\omega\varepsilon k_r J_l'(k_r r) \\ 0 \end{pmatrix} e^{il\varphi} e^{-k_z z} \quad (5)$$

Subsequently, the spin angular momentum (SAM) of the e-OV can be calculated by

$$\langle \boldsymbol{S} \rangle = \frac{1}{2\omega} \operatorname{Im}\left[ \varepsilon \boldsymbol{E}^* \times \boldsymbol{E} + \mu \boldsymbol{H}^* \times \boldsymbol{H} \right]. \quad (6)$$

Substituting Eq. (5) into Eq. (6), one can obtain

$$\langle \boldsymbol{S} \rangle = \begin{pmatrix} \langle S_r \rangle \\ \langle S_\varphi \rangle \\ \langle S_z \rangle \end{pmatrix} = \begin{pmatrix} A^2 \dfrac{\varepsilon k_z k_r^2 l}{2\omega r} J_l(k_r r)^2 e^{-2k_z z} \\ 0 \\ A^2 \dfrac{\varepsilon k_r^3 l}{2\omega r} J_l'(k_r r) J_l(k_r r) e^{-2k_z z} \end{pmatrix} \quad (7)$$

With each component of the SAM, we can plot the spin structure of the e-OV (Fig. 1 in the main text), which is a direct photonic analogue of a Neel-type magnetic skyrmion [18].

In order to further verify the skyrmion analogy, we calculate the skyrmion number (*n*) of the spin structure [15]:

$$n = \frac{1}{4\pi} \iint \mathbf{M} \cdot \left( \frac{\partial \mathbf{M}}{\partial x} \times \frac{\partial \mathbf{M}}{\partial y} \right) dxdy \qquad (8)$$

where **M** stands for the unit vector in the direction of the local SAM vector within the e-OV, and the integral is taken over a unit cell of skyrmion in the horizontal plane. Since the SAM contains only the $r$- and $z$-component, **M** can be represented as

$$\mathbf{M} = \begin{pmatrix} M_r \\ M_\varphi \\ M_z \end{pmatrix} = \begin{pmatrix} \cos\theta_r \\ 0 \\ \sin\theta_r \end{pmatrix}, \qquad (9)$$

where $\vartheta_r$ is the orientation angle of the SAM vector with respect to the X-Y plane, and independent of the azimuthal angle $\varphi$. Thus, Eq. (8) results in

$$n = \frac{1}{2} \int_0^{r_0} \cos\theta_r \frac{\partial \theta_r}{\partial r} dr . \qquad (10)$$

Here, $r_0$ is the position where the spin vector turns in opposite direction to the central position $r=0$. This radius $r_0$ defines a unit cell of skyrmion in the geometrical central area. For a spin structure with the SAM changing progressively from the "up" state in the centre $r=0$ to the "down" state at $r=r_0$ (Fig. 1b, top panel), Eq. (10) can be evaluated as n = 1. Thus, the skyrmion number of an e-OV has an integer value of "1" or "-1", with the sign which depends on the chiral property (positive or negative spin is observed at $r=0$) of e-OV.

The skyrmion number is related to the topological property of an electromagnetic field. In Fig. S1 (a-b), we show the intensity profiles of an e-OV ($L=1$) in the X-Y and X-Z planes. The intensity of an e-OV has a donut-shaped profile, which has genus 1. This results in a photonic skyrmion with an integer number of 1. Indeed, each standing wave fringe in an e-OV can be considered as a well-defined donut structure. As a result, a photonic skyrmion lattice is indeed formed along the radial direction, with the SAM vector distributions shown in Fig. S1 (c,d).

## 4. Theoretical analysis of photonic spin structure in a propagating vortex beam.

In the case of a propagating beam with an in-plane wave-vector $k_r<k$, the electric and magnetic fields of a TM-mode can be calculated by substituting Eq. (4) into Eq. (1):

$$\boldsymbol{E}(k_r) = \begin{pmatrix} E_r \\ E_\varphi \\ E_z \end{pmatrix} = \begin{pmatrix} iAk_z k_r J_l{}'(k_r r) \\ -A\dfrac{lk_z}{r} J_l(k_r r) \\ Ak_r^2 J_l(k_r r) \end{pmatrix} e^{il\varphi} e^{ik_z z}$$

$$\boldsymbol{H}(k_r) = \begin{pmatrix} H_r \\ H_\varphi \\ H_z \end{pmatrix} = \begin{pmatrix} A\dfrac{\omega\varepsilon l}{r} J_l(k_r r) \\ Ai\omega\varepsilon k_r J_l{}'(k_r r) \\ 0 \end{pmatrix} e^{il\varphi} e^{ik_z z} \qquad (11)$$

This describes an ideal diffraction-free Bessel vortex beam propagating in z-direction. Subsequently, the SAM can be calculated by Eq. (6) as

$$\langle \boldsymbol{S} \rangle = \begin{pmatrix} \langle S_r \rangle \\ \langle S_\varphi \rangle \\ \langle S_z \rangle \end{pmatrix} = \begin{pmatrix} 0 \\ A^2 \dfrac{\varepsilon k_z k_r^3}{2\omega} J_l{}'(k_r r) J_l(k_r r) \\ A^2 \dfrac{\varepsilon k_r l}{2\omega r}\left(k_z^2 + \varepsilon_r k_0^2\right) J_l{}'(k_r r) J_l(k_r r) \end{pmatrix} \qquad (12)$$

As can be seen, in contrast to an e-OV for which the SAM contains only the *r*- and *z*-components and forms a Neel-type skyrmion, the SAM of a propagating Bessel vortex beam contains only the *φ*- and *z*-components (Fig. S2 (a)) which is characteristic for a Bloch-type skyrmion. The orientation angle ($\vartheta_\varphi$) of the SAM vector with respect to the X-Y plane obtained from

$$\tan\theta_\varphi = \frac{S_z}{S_\varphi} \qquad (13)$$

is plotted in Fig. S2 (b) together with the SAM vector distribution in the *x*-direction (Fig. S2 (c)). As can be seen, although the SAM vector changes progressively towards the azimuthal direction at the central area, there is an abrupt change of SAM at the positions where $S_\varphi=0$, and the SAM vector will never become anti-parallel to the central one, so that skyrmion is not formed. The reason is that a diffraction-free propagating beam is boundary-free in the propagation direction, thus unable to form a well-defined topology structure. This is in contrast to an e-OV with the intensity decaying exponentially in *z*-direction because of the evanescent nature of the field.

However, in a focused vortex beam (FVB), this abrupt change of SAM can be overcome, enabling us to achieve a skyrmion-like photonic spin structure of Bloch type in the central area. In a FVB, the diffracted field near focal plane can be evaluated using the Debye approximation, with a rigorous theory developed by Richards and Wolf [23]. To achieve a TM-mode near focal plane, a radially-polarized optical vortex (topological charge $l$) is used as incident beam. After focusing, the electromagnetic field near focal plane can be calculated with the Richards-Wolf theory as:

$$\boldsymbol{E} = \begin{pmatrix} E_r \\ E_\varphi \\ E_z \end{pmatrix} = \begin{pmatrix} \int_0^{\theta_{max}} A(\theta)\{ik_z J'_l(k_r r)e^{il\varphi}\}e^{ik_z z}d\theta \\ \int_0^{\theta_{max}} A(\theta)\{-\frac{lk_z}{k_r r}J_l(k_r r)e^{il\varphi}\}e^{ik_z z}d\theta \\ \int_0^{\theta_{max}} A(\theta)\{k_r J_l(k_r r)e^{il\varphi}\}e^{ik_z z}d\theta \end{pmatrix} = \int_0^{\theta_{max}} A(\theta)\frac{1}{k_r}\boldsymbol{E}(k_r)d\theta$$

$$\boldsymbol{H} = \begin{pmatrix} H_r \\ H_\varphi \\ H_z \end{pmatrix} = \begin{pmatrix} \int_0^{\theta_{max}} A(\theta)\{\frac{\omega\varepsilon l}{k_r r}J_l(k_r r)e^{il\varphi}\}e^{ik_z z}d\theta \\ \int_0^{\theta_{max}} A(\theta)\{i\omega\varepsilon J'_l(k_r r)e^{il\varphi}\}e^{ik_z z}d\theta \\ 0 \end{pmatrix} = \int_0^{\theta_{max}} A(\theta)\frac{1}{k_r}\boldsymbol{H}(k_r)d\theta$$

(14)

with

$$A(\theta) = i^{m+1} A_r t^p(\theta) f e^{-ikf} l_0(\theta)\sqrt{\frac{n_0}{n_1}\cos\theta}\sin\theta \, , \quad (15)$$

where $f$ is the focal length, $t^p(\vartheta)$ is the transmission coefficient of the multilayer structure located at the focal plane for a $p$-polarized light, $l_0(\vartheta)$ is the amplitude profile of incident beam, $\vartheta_{max}$ is the maximum incident angle allowed by the focusing lens which is determined by the numerical aperture (NA) as NA= $n_0 \cdot \sin(\vartheta_{max})$, $n_0$ and $n_1$ are the refractive indices of media in the incident and output space, respectively. The in-plane wave-vector ($k_r$) is related to the incident angle ($\vartheta$) by $k_r = n_0 k_0 \sin\vartheta$. As the result, the electromagnetic field near a focal plane is the superposition of the diffraction-free Bessel vortex beams with different in-plane wave-vectors (Eq. (14)).

After obtaining the electromagnetic field, the SAM of a FVB near a focal plane (Fig. S2 (d)) can be calculated using Eq. (6). Comparing the SAM distribution and the orientation angle of the SAM vector for ideal and focused vortex beams (cf Fig. S2 (a-c) and Fig. S2 (d-f)), we can find a remarkable difference. The $S_z$ and $S_\varphi$ SAM components do not intersect at the zero point for a FVB as they do for a nondiffracting vortex beam, and there exists a spin state at $r=r_0$ where $S_\varphi=0$ and the SAM vector (<$S$>= $S_z$) is anti-parallel to one at the centre: the spin vectors which point to the $+z$ direction at the geometric centre, change progressively towards the azimuthal direction, and eventually to the opposite $-z$ direction at $r=r_0$. As the result, a Bloch-type skyrmion-like photonic spin structure is formed within the central circular area that is defined by $r_0$ [18]. Nevertheless, it must be mentioned that outside this area, the spin vector does not exhibit a regular change from the positive to the negative state any more. As the result, this cannot be treated as a rigorous analogue to a Bloch-type skyrmion.

## 5. Numerical investigation of the optical response of a polystyrene nanosphere.

In this work, polystyrene nanosphere with diameter of 320 nm is employed as the near-field scatter in order to out-couple the SPP field with a certain local spin state into the far-field scattering radiation of elliptical polarization. In order to understand the optical response of the PS nanosphere, we performed numerical simulation with finite-difference time-domain method by using the commercial software Lumerical.

The schematic diagram of the structure is shown in Fig. S6 (a). In the simulation, the silver film thickness is set to 50 nm. The refractive indices of air, glass substrate and PS nanosphere are set to 1, 1.51 and 1.48, respectively, and that for the silver film is obtained from the Material Database in the software. The gap between the nanosphere and silver film is set to 2 nm, to mimic the separation that created by the 4-MBA molecules, which is employed in the experiment to facilitate the formation of isolated PS nanospheres on the silver film. In the simulation set-up, the incident beam at 633 nm with total field scattered field (TFSF) scheme is employed.

Figure S6 (b-c) shows the calculated far-field scattering radiation patterns from the PS nanosphere at the Fourier domain, under the illumination with an *x*-polarized and a *z*-polarized incident light, respectively. As can be seen, a transversal electric field excites a scattering radiation mainly concentrated at small angles while a longitudinal electric field primary at large angles. Considering the NA of the collection objective lens that is used in the experiment (NA=0.7), the collection efficiency ratio is plotted in Fig. S6 (d) for various nanosphere sizes. The plot shows that the PS nanosphere is more sensitive to the transverse electric field component of the evanescent wave than to the longitudinal one. Their collection efficiency ratio is optimized for a diameter of approximately 300 nm for NA = 0.7, showing more than 2 orders of magnitude rejection. This ensures that our set-up characterizes the spin property that is associated with the transverse field of SPP, despite a strongly dominant longitudinal electric field in SPP wave (typically, the intensity of longitudinal field is 1-order of magnitude higher than the transverse one in the visible spectral range).

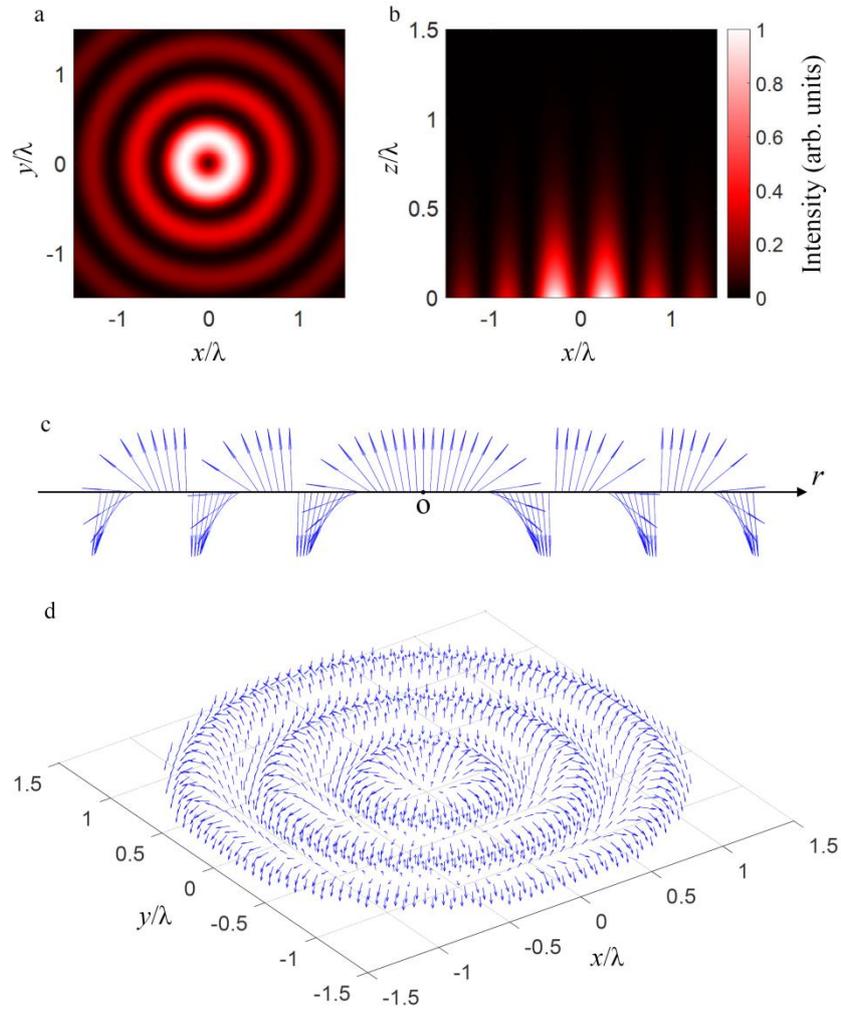

Figure S1. Structure of a Neel-type photonic skyrmion. (a,b) Normalized intensity profile of an e-OV of topological charge $L=1$ in (a) x-y and (b) x-z planes for $k_r=1.05k$. (c,d) The corresponding SAM vector distributions of the e-OV in (c) radial direction and (d) x-y plane. The axes in (a), (b) and (d) are in the unit of wavelength.

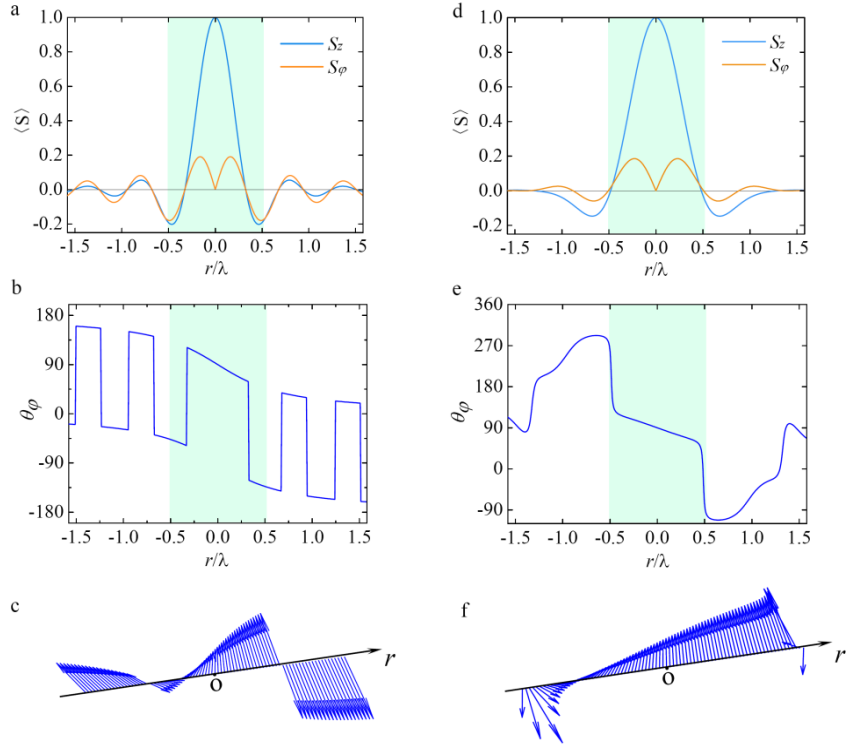

Figure S2. Bloch-type skyrmion-like photonic spin structure in a focused vortex beam. (a) The normalized SAM of an ideal diffraction-free Bessel vortex beam of topological charge $l=1$ with $k_r=0.9k$ containing only $\varphi$- and $z$-components. (b) The orientation angle of the SAM vector with respect to the x-y plane and (c) the SAM vector distribution along the *x*-axis (within the light green region in (b)) calculated from the data in (a). (d-f) The same as in (a-c) for a focused vortex beam obtained from a radially-polarized optical vortex beam with numerical aperture NA=0.9, which a superposition of (a) with different $k_r$ wavevectors (0 - $0.9k_0$) determined by NA.

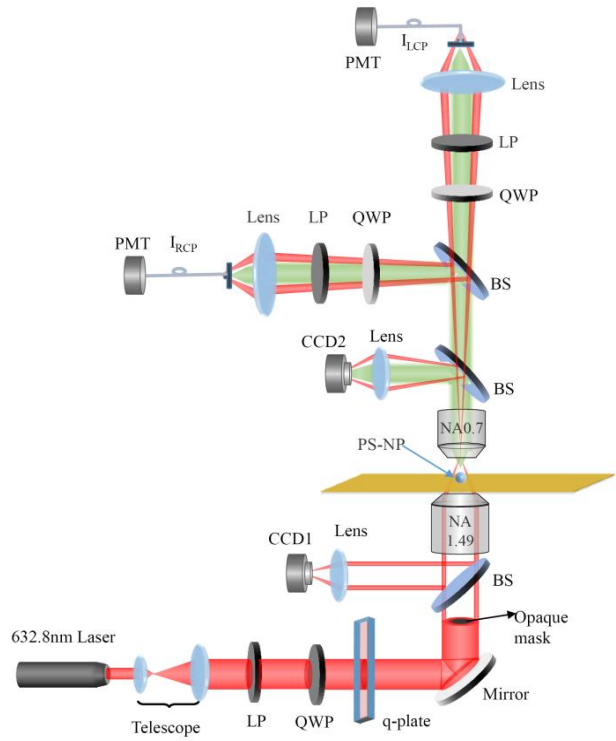

Figure S3. Experimental setup for characterization of the spin structure in a SPP vortex. LP: linear polarizer; HWP: half wave plate; QWP: quarter wave plate; BS: beam splitter; PMT: photo-multiplier tube; CCD: charge-coupled device.

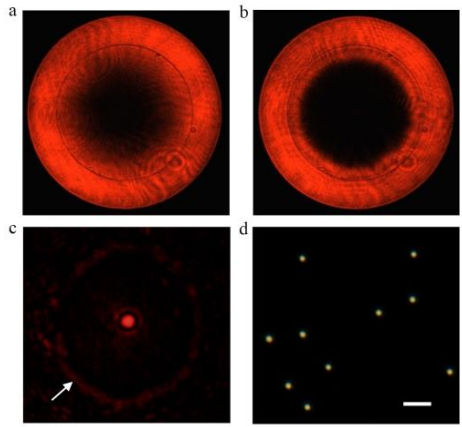

Figure S4. (a) Back focal plane image of the reflected beam from the oil-immersion objective lens, where a narrow dark ring indicates the excitation of SPP at the air-silver interface. (b) The same image when a circular opaque mask is inserted into the beam path, which blocks incident light at low angles while does not affect the excitation of the SPPs. (c) Rayleigh scattering image of a PS nanoparticle captured with CCD2 from the transmission side, clearly illustrating the spatial separation of the scattering signal from the PS nanosphere (central bright spot) and the directly transmitted light (the outside ring profile marked with a white arrow. (d) Dark-field image of the isolated PS nanospheres immobilized on the silver film. The scale bar in (d) is 5 μm.

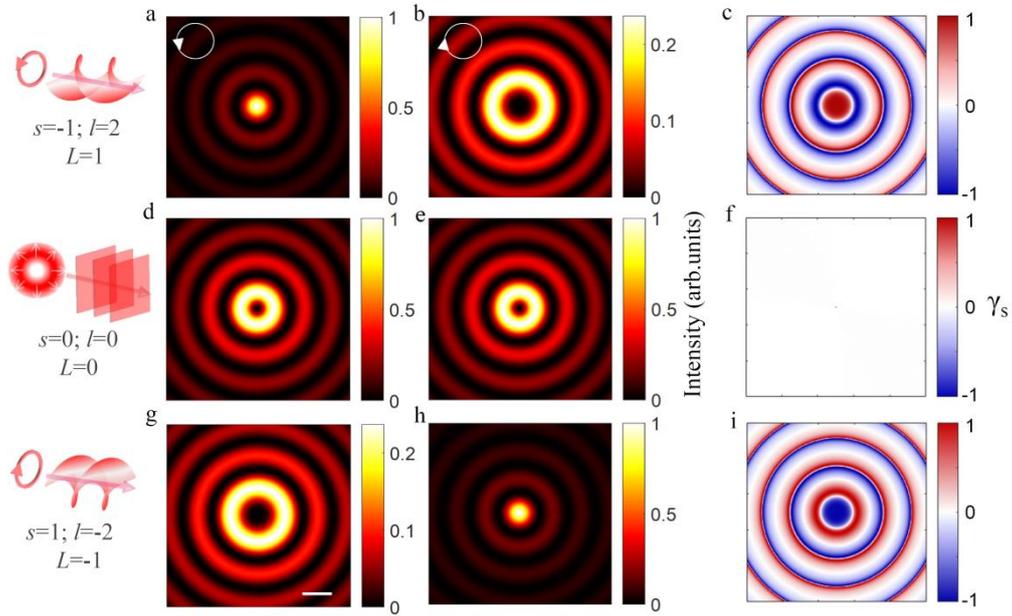

Figure S5. Simulated spin-structure of the plasmonic vortex when $k_r=k_{spp}=1.05k_0$, for the purpose of comparison with the experimental results in Fig. 2 in the main text. All images are of the same size; the scale bar in (g) is $\lambda/2$.

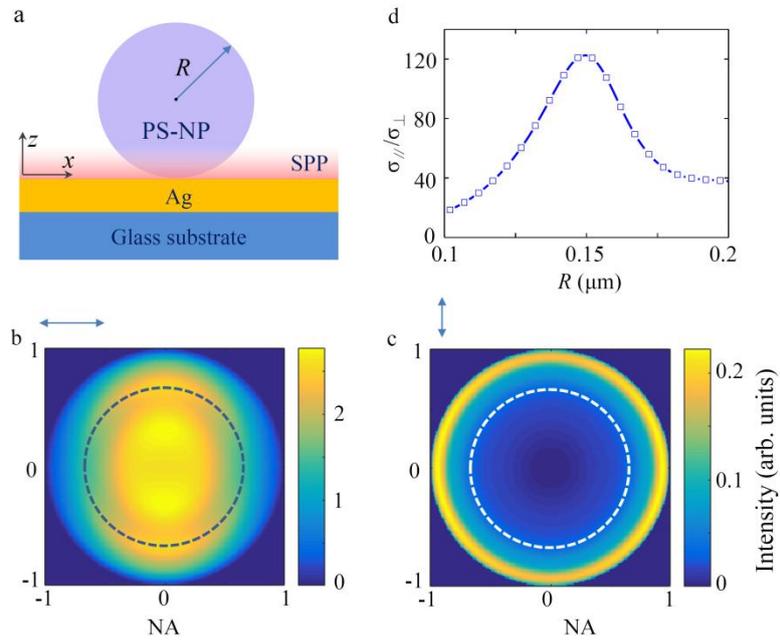

Figure S6. (a) Schematic diagram of the structure investigated in the simulation. (b-c) The calculated far-field scattering radiation patterns from the PS nanosphere in the Fourier domain, under the illumination with an *x*-polarized and a *z*-polarized incident light, respectively. The dashed circle represents the NA of the collection system employed in the experiment. (d) The ratio of the collection efficiencies of the scattering of the transverse and longitudinal electric fields for NA=0.7.